\documentclass[
reprint,
showpacs,preprintnumbers,
amsmath,amssymb,
aps,
pre,
]{revtex4-1}

\usepackage{acronym}
\usepackage{amsmath}
\usepackage{amssymb}
\usepackage{color}
\usepackage{graphicx}
\usepackage{grffile}
\usepackage{hyperref}
\usepackage{ifthen}
\usepackage[utf8]{inputenc}
\usepackage{upgreek}
\usepackage{url}
\usepackage[dvipsnames]{xcolor}
\usepackage{xspace}
\usepackage{rviewport}

\newcommand\eg{e.\,g.,\ }
\newcommand\ie{i.\,e.,\ }
\newcommand\cf{cf.\ }
\newcommand\bracket[1]{[#1]}

\newcommand\fig[1]{Fig.~\ref{#1}}

\newcommand\figur[1]{Figure~\ref{#1}}

\newcommand\subfig[2]{Fig.~\ref{#1}\sub{#2}}
\newcommand\subfigs[2]{Figs.~\ref{#1}\sub{#2}}
\newcommand\subfigure[2]{Figure~\ref{#1}\sub{#2}}
\newcommand\subfigures[2]{Figures~\ref{#1}\sub{#2}}
\newcommand\sub[1]{(#1)}
\newcommand\Eq[1]{Eq.~\eqref{#1}}
\newcommand\Eqs[1]{Eqs.~\eqref{#1}}

\newcommand{\quarternl}{0.23851\textwidth}
\newcommand{\thirdnl}{0.32978\textwidth}
\newcommand{\halfnl}{0.478\textwidth}
\newlength{\snapwidth}

\renewcommand\vec[1]{\mathbf{#1}}

\newcommand\fromto[2]{{\ensuremath{A_{#1}\!\!\rightarrow\!\! A_{#2}}}}
\newcommand\pwall{\ensuremath{P_\mathrm{wall}}\xspace}
\newcommand\pint{\ensuremath{P_\mathrm{int}}\xspace}
\newcommand{\sft}{\ensuremath{s_t}\xspace}
\newcommand{\sfo}{\ensuremath{s_Y}\xspace}
\newcommand{\ex}{_\mathrm{E}}
\newcommand{\theo}{_\mathrm{D}}
\newcommand{\init}{_\mathrm{I}}
\newcommand{\fin}{_\mathrm{F}}

\newacro{2d}[2D]{two-dimensional}

\begin{document}
	
	\title{Ballistic propagation of density correlations and excess wall forces in quenched granular media}
	
	\author{Thomas Schindler}
	\email{thomas.schindler@fau.de}
	\affiliation{Theoretische Physik 1, FAU Erlangen-N\"urnberg, Staudtstr.~7, 91058 Erlangen, Germany}
	\author{Christian M.\ Rohwer}
	\email{christian.rohwer@uct.ac.za}
	\affiliation{Department of Mathematics and Applied Mathematics, University of Cape Town, 7701 Rondebosch, Cape Town, South Africa}
	\affiliation{Max Planck Institute for Intelligent Systems, Heisenbergstr.~3, 70569 Stuttgart, Germany}
	\affiliation{4th Institute for Theoretical Physics, University of Stuttgart, Pfaffenwaldring 57, 70569 Stuttgart, Germany}
	\date{\today}
	
	\begin{abstract}
		We investigate a granular gas in a shaken quasi-two-dimensional box in molecular dynamics computer simulations.
		After a sudden change (quench) of the shaking amplitude, transient density correlations are observed orders of magnitude beyond the steady-state correlation length scale. 
		Propagation of the correlations is ballistic, in contrast to recently investigated quenches of Brownian particles that show diffusive propagation
		[Rohwer \textit{et al.}, Phys.\ Rev.\ Lett., \textbf{118}, 015702 (2017),
		Rohwer \textit{et al.}, Phys.\ Rev.\ E, \textbf{97}, 032125 (2018)].
		At sufficiently strong cooling of the fluid the effect is overlaid by clustering instability of the homogeneous cooling state with different scaling behavior.
		We are able to identify different quench regimes. In each regime correlations exhibit remarkably universal position dependence.
		In simulations performed with side walls we find confinement effects for temperature and pressure in steady-state simulations, and an additional transient wall pressure contribution upon changing the shaking amplitude.
		The transient contribution is ascribed to enhanced relaxation of the fluid in the presence of walls.
		From incompatible scaling behavior we conclude that the observed effects with and without side walls constitute distinct phenomena.
	\end{abstract}
	
	\maketitle

	\section{Introduction}\label{sec:intro}
	
	A dynamic system of macroscopic particles tends to dissipate kinetic energy due to inelastic collisions.
	In order to maintain particle motion, energy input by an external source is needed.
	One of the setups commonly employed to this end is the quasi-\ac{2d} granular shaker, which consists of a flat box filled with typically millimeter-sized beads (usually made of metal or glass), that is vibrated vertically.
	The directed energy input is randomized in particle-particle collisions yielding dynamical steady states reminiscent of thermal equilibrium.
	In particular, the parameter space for formation of regular lattices, fluids, and coexistence thereof~\cite{prevost2004nonequilibrium,melby2005dynamics,reis2006crystallization,clerc2008liquid,vega2008effect,rivas2011segregation,guzman2018critical,schindler2019nonequilibrium}
	bears analogy to the corresponding \ac{2d} equilibrium system~\cite{schmidt1997phase}.
	However, several properties reveal the nonequilibrium nature of the steady states,
	such as
	inelastic collapse at the bottom of the container~\cite{olafsen1998clustering,nie2000dynamics,olafsen2005two,khain2011hydrodynamics},
	inhomogeneous granular temperatures~\cite{prevost2004nonequilibrium,lobkovsky2009effects},
	non-Gaussian velocity distributions~\cite{losert1999velocity,olafsen1999velocity,kawarada2004non},
	segregation of mixtures~\cite{rivas2011sudden,rivas2011segregation,rivas2012characterization},
	and inelastic hydrodynamic modes~\cite{brito2013hydrodynamic}.
	Additionally, granular systems are known to exhibit nonequilibrium collective phenomena such as flocking~\cite{kumar2014flocking} and pattern formation~\cite{aranson2006patterns}.
	
	In the present study we disturb the steady state by changing the driving strength, in order to search for further evidence of its nonequilibrium origin.
	This technique has proven fruitful as several anomalies in response functions have been reported,
	\eg in the Kovacs memory effect~\cite{prados2014kovacs,trizac2014memory,brey2014memory}
	or in the compaction behavior~\cite{nicodemi1999dynamical,caglioti1997tetris,brey2001linear,brey2002memory}. Our concrete goal is to illustrate the emergence of collective phenomena (\ie transient correlations on large scales) that reveal how nonequilibrium states of driving and dissipation are fundamentally different from thermal equilibrium--a fact that is disguised by the phenomenological resemblance of the steady states.
	We go about this by analyzing the structure of correlations emerging after a change in the vibration amplitude.
	Additionally, we study the differences to Brownian dynamics following quenches, in which similar correlations have been reported~\cite{rohwer2017transient,rohwer2018nonequilibrium}, in order to uncover the origin of the observed effects.
	
	Part of the article is devoted to classical Casimir forces~\cite{kardar1999friction} between distant walls, mediated by the granular medium.
	The term is borrowed from the corresponding quantum effect~\cite{bordag2009advances} related to fluctuating electromagnetic fields confined in a two-plate geometry (Casimir geometry).
	When the properties of the medium are altered by the confinement, either through a modification of the fluctuation modes of the medium (fluctuation induced) or by an alteration of the fluid density (density induced), this can result in nontrivial macroscopic forces between the confining surfaces~\cite{rodriguez2016clustering,aumaitre2001segregation,villanueva2010casimir,denisov2011simulation,zuriguel2005role}.
	Both effects have been demonstrated in the Brownian reference system. The question addressed in this article is whether the shaken granulates also exhibit such forces and whether these can be classified as either fluctuation induced or density induced.
	
	The paper is organized as follows.
	In Sec.~\ref{sec:model} we introduce the examined setup and give technical details about simulation parameters.
	Our results for density correlations in bulk simulations are presented in Sec.~\ref{sec:quench-bulk}.
	We observe large-scale correlations after a change in driving amplitude, for which we carve out similarities and differences to Brownian dynamics.
	Section~\ref{sec:casimir} treats temperature and pressure in a geometry with additional side walls, 
	and discusses their finite-size scaling in steady state and after quenches.
	In Sec.~\ref{sec:con} we conclude by classifying the observed phenomena.

	\section{Setup and model}\label{sec:model}
	
	\begin{figure}
		\begin{center}
			\includegraphics[width=\halfnl]{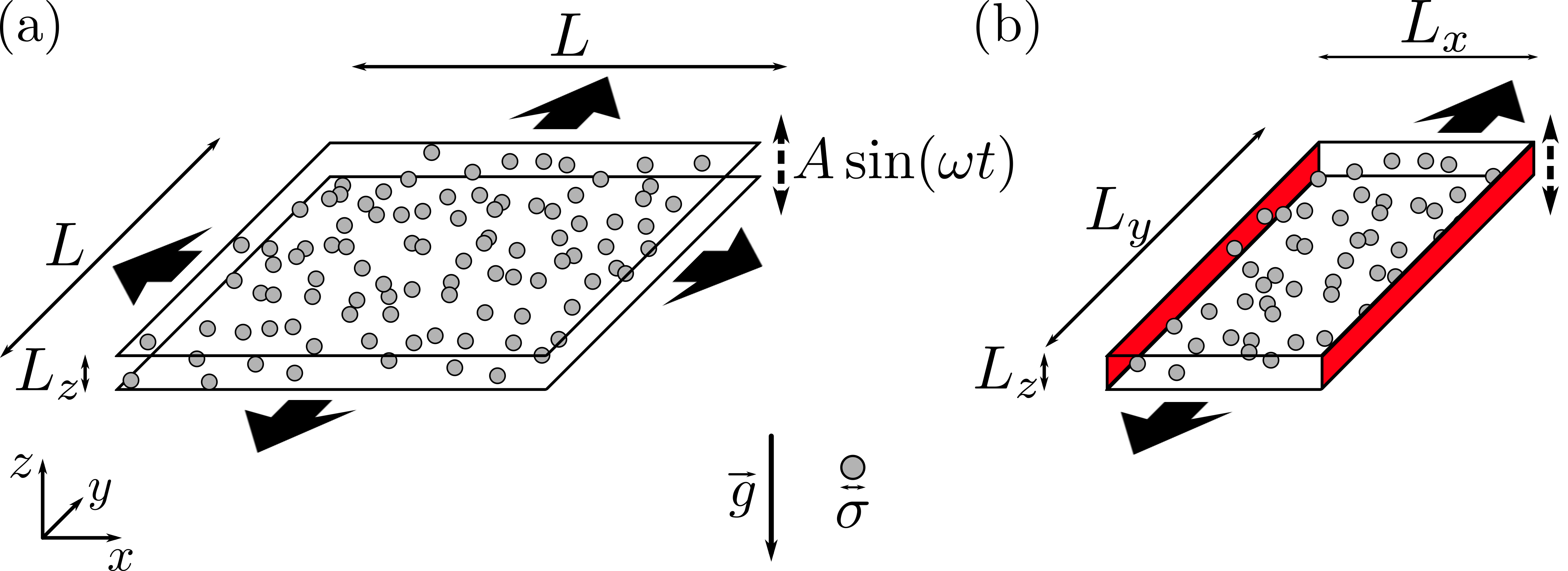}
		\end{center}
		\caption{\label{fig:set}%
			Sketches of the setup \sub{a} in bulk geometry and \sub{b} with side walls (red) confining the simulation box in the $x$ direction.
			The bold arrows indicate the directions in which periodic boundary conditions are applied.
			Dashed arrows mark the shaking in $z$ direction.
			Symbols are declared in the main text.
		}
	\end{figure}
	
	We consider a system of $N$ hard spheres of diameter $\sigma$ and mass $m_\mathrm{p}$ in a shallow cuboidal box with dimensions $L_x\times L_y\times L_z$ ($L_z=2\sigma$) with hard bottom and top plates (see \fig{fig:set}).
	Gravitational acceleration $g$ acts in the negative $z$ direction, which induces a time scale $\tau_0\equiv\sqrt{\sigma/g}$ and an energy scale $\epsilon\equiv m_\mathrm{p}g\sigma$.
	The two plates are oscillating in phase with a time-dependent displacement in the $z$ direction and their vertical positions are described by $\pm L_z/2+A\sin(\omega t + \varphi)$ with an amplitude $A$, an
	angular frequency $\omega=50\tau_0^{-1}$, time $t$, and a phase shift $\varphi$.
	The sole purpose of introducing $\varphi$ here is to clarify that quenches (explained below) are not in sync with the plate oscillation;
	this is achieved by averaging over $\varphi$.
	The area density of particles is fixed to $\rho\equiv N/L_x L_y=0.5\sigma^{-2}$ throughout the paper.
	
	The main control parameter in this article is $A$, and three different amplitudes, $A_1\equiv 0.002\sigma$, $A_2\equiv 0.005\sigma$, and $A_3\equiv 0.05\sigma$, are considered.
	We conduct steady-state simulations at constant $A$, as well as quenched simulations.
	Quenches are performed by suddenly increasing or decreasing the shaking amplitude at $t=0$ from an initial value $A\init$ to a final value $A\fin$ and observing the granular fluid after the quench.
	Henceforth we denote quench protocols between $A_1$ and $A_2$ as ``moderate quenches'' and protocols starting from or ending at $A_3$ (with $A_3$ being an order of magnitude larger than $A_1$ and $A_2$) as ``strong quenches''.
	As the setup is very shallow, we treat it as an effective \ac{2d} system and calculate observables only from the $x$ and $y$ components of particle positions and velocities.
	
	Two different geometries are investigated.
	On the one hand, we employ the described setup with periodic boundary conditions in the $x$ and $y$ directions \bracket{see \subfig{fig:set}{a}}, henceforth referred to as (\ac{2d}-)bulk.
	The lateral box dimensions here are square shaped, with $L\equiv L_x=L_y$ ranging from $100\sigma$ to $400\sigma$.
	The other setup considered has periodic boundary conditions in the $y$ direction (with $L_y=400\sigma$) but is confined between two vertical immovable side walls in the $x$ direction, separated by a distance $L_x$ ranging from $5\sigma$ to $200\sigma$ \bracket{\subfig{fig:set}{b}}.
	Beyond the side walls we assume an additional infinitely extended exterior domain (not shown) with the same filling density, as would be present in the well-known Casimir setup of two planar walls at finite distance immersed in bulk.
	The exterior cannot be simulated explicitly due to its infinite size.
	Instead, we extrapolate simulation results of the interior (obtained for the different $L_x$) to $\infty$ in order to emulate the exterior.
	The main goal here is the measurement of finite-size deviations (from the limit $L_x\rightarrow\infty$) in temperature and pressure as differences between the actually simulated interior region and the extrapolated exterior.
	
	The setup is studied in event driven molecular dynamics computer simulations with the DynamO~\cite{bannerman2011dynamo} package.
	The central idea of the algorithm is to predict collisions of particles from their current positions and velocities.
	These are entered into a schedule.
	The system is then evolved by forwarding to the next collision in the schedule and calculating the new velocities and the next collisions of the collision partners.
	The algorithm is suitable for a system with short interaction times and parabolic trajectories in between, such as the hard macroscopic marbles studied here.
	The interactions are modeled as instantaneous billiard-like collisions with momentum conservation~\cite{rapaport2004art}.
	Energy loss is accounted for by rescaling the relative particle velocities after collisions (either with other particles or with walls) by a coefficient of restitution of 0.95.
	Coulomb friction (\ie friction due to relative tangential motion at contact) is neglected, and hence there is no transfer of angular momentum.
	Therefore, we do not need to simulate rotations of the spheres.
	While this model is simplistic, it captures the essential mechanisms of energy input and dissipation, and thus creates the nonequilibrium steady states that are also found in experiments or more sophisticated simulations.
	In our previous studies~\cite{schindler2019nonequilibrium}
	we found the phase behavior to be consistent with simulations employing rotating spheres~\cite{prevost2004nonequilibrium,melby2005dynamics,reis2006crystallization,clerc2008liquid,vega2008effect,rivas2011segregation,guzman2018critical}.

	\section{Quenched bulk}\label{sec:quench-bulk}
	
	In this section we characterize the bulk system \bracket{see \subfig{fig:set}{a}} after a quench and show how large-scale transient correlations emerge.
	
	\begin{figure}
		\includegraphics[width=\halfnl]{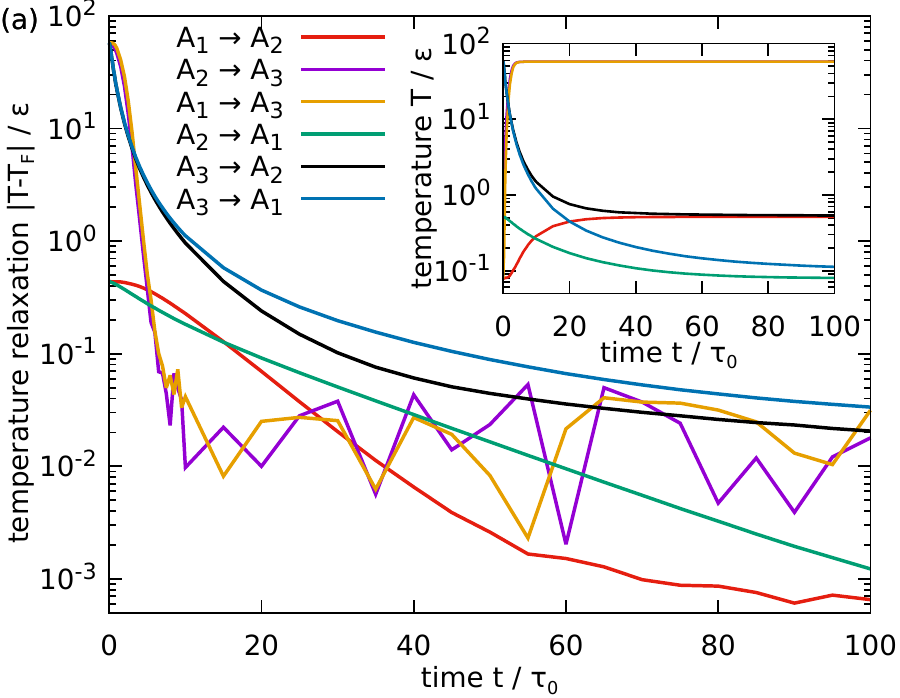}
		\includegraphics[width=\halfnl]{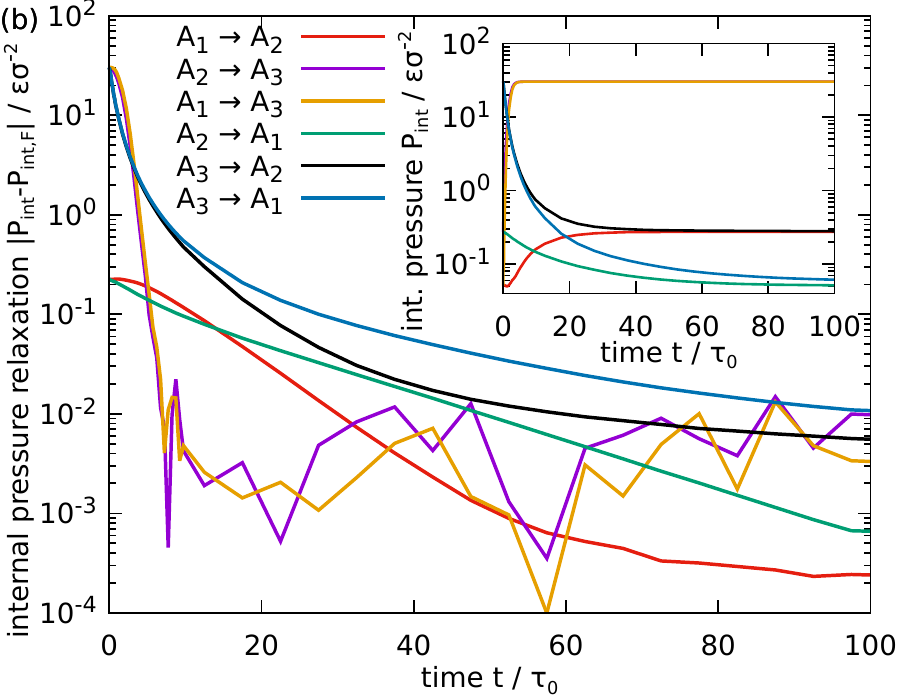}
		\caption{\label{fig:PT-bqA}%
			\sub{a} Relaxation of the temperature $T$ towards the final steady-state value $T\fin$ in the quenched fluid as a function of time $t$ after the quench on logarithmic scale for several initial and final amplitudes $A_i$ ($i=1,2,3$) as indicated.
			The insets show the values of $T$ itself.
			The $T\fin$ at each $A_i$ are obtained in separate steady-state simulations.
			Panel \sub{b} shows the same analysis as \sub{a} but for the internal pressure $\pint$.
			Data for $A\fin=A_3$ (orange and purple lines) at $t>7\tau_0$ display noise which is proportional to the higher values of $T\fin$ or $P_\mathrm{int,F}$, respectively.
		}
	\end{figure}
	
	\subsection{Global observables and density inhomogeneity}\label{sec:quench-bulk-global}
	
	At first, however, we look at the
	granular temperature and internal mechanical pressure, defined as
	\begin{equation}
		\label{eq:Temp}
		T\equiv\left\langle \frac{1}{Nd}\sum_{i=1}^{N}\frac{m_\mathrm{p}}{2}\vec{v}_i^2\right\rangle
	\end{equation}
	and
	\begin{equation}
		\label{eq:Pint}
		\pint\equiv\left\langle\frac{1}{d L_x L_y\Delta t}\sum_{\mathrm{PP}}\Delta\vec{p}_i\cdot(\vec{r}_i-\vec{r}_j)\right\rangle\,,
	\end{equation}
	respectively. 
	Here $\vec {v}_i$ is the velocity of particle $i$, $d=2$ is the spatial dimension,
	$\mathrm{PP}$ indicates summation over all particle-particle collisions between particles $i$ and $j$ during a time interval $\Delta t$, $\Delta\vec{p}_i$ is the change of momentum of particle $i$ during the collision, and $\vec{r}_i$ is its position.
	Angular brackets in both equations denote averages over a large time interval in steady-state simulations or over a small time interval $[t-\Delta t/2,t+\Delta t/2]$ and multiple quench realizations in quenched simulations.
	We stress once more that all vectors in the above equations are \ac{2d} projections onto the $xy$ plane.
	
	Figure~\ref{fig:PT-bqA} shows the two quantities defined in \Eqs{eq:Temp} and \eqref{eq:Pint} as functions of time $t$ after the quench for several $A\init$ and $A\fin$.
	We observe two qualitatively different types of behaviors.
	In the case of heating or moderate cooling (\fromto21), $T$ and $\pint$ relax exponentially in time towards the final steady-state values $T\fin$ and $P_\mathrm{int,F}$, respectively.
	The relaxation time is inversely proportional to $A\fin$,
	since $A\fin$ is proportional to the root mean square particle velocity.
	This proportionality is not necessarily true in general but it does apply in the strong shaking regime ($A\omega^2\gg g$) employed in this work.
	Here $\omega$ provides the predominant timescale (see, \eg Refs.~\cite{rivas2011segregation,melby2005dynamics}) and therefore the particle velocities scale with the peak velocity of the plates, $A\omega$.
	Heating protocols exhibit an additional start-up phase of slow temperature and pressure change.
	The reason for this is the two-step process in which energy is transferred in particle-plate collisions to the vertical degrees of freedom at first and to the horizontal degrees of freedom only in secondary particle-particle collisions.
	In contrast, cooling protocols lack the start-up phase, as dissipation takes place homogeneously in every collision.
	
	Only in the case of strong cooling (\ie with initial amplitude $A_3$) do $T$ and $\pint$ relax algebraically in time.
	This behavior is known for a freely cooling granular gas~\cite{haff1983grain} which is comparable to the present situation as long as $T(t)\gg T\fin$.
	
	The asymmetry between strong heating and strong cooling originates from the asymmetry of energy gain and dissipation of the horizontal degrees of freedom via particle-particle collisions.
	Cooling through inelastic dissipation takes place in every collision.
	Heating of the horizontal directions, on the other hand, only occurs, if a particle has been accelerated by the oscillating plates and then transfers its energy to horizontal motion in a particle-particle collision as has been demonstrated for steady-state fluctuations in a setup with particles of differing masses~\cite{rivas2011sudden,rivas2012characterization}.
	Moderate quenches constitute only weak disturbances of the steady state where the described asymmetry plays only a minor role.
	As we shall see, the two described cases are distinct by other observables as well.
	Hence, we will refer to heating or moderate cooling as type~I and to strong cooling as type~II behavior in the remainder of the paper.

	\begin{figure*}
		\includegraphics[width=\textwidth]{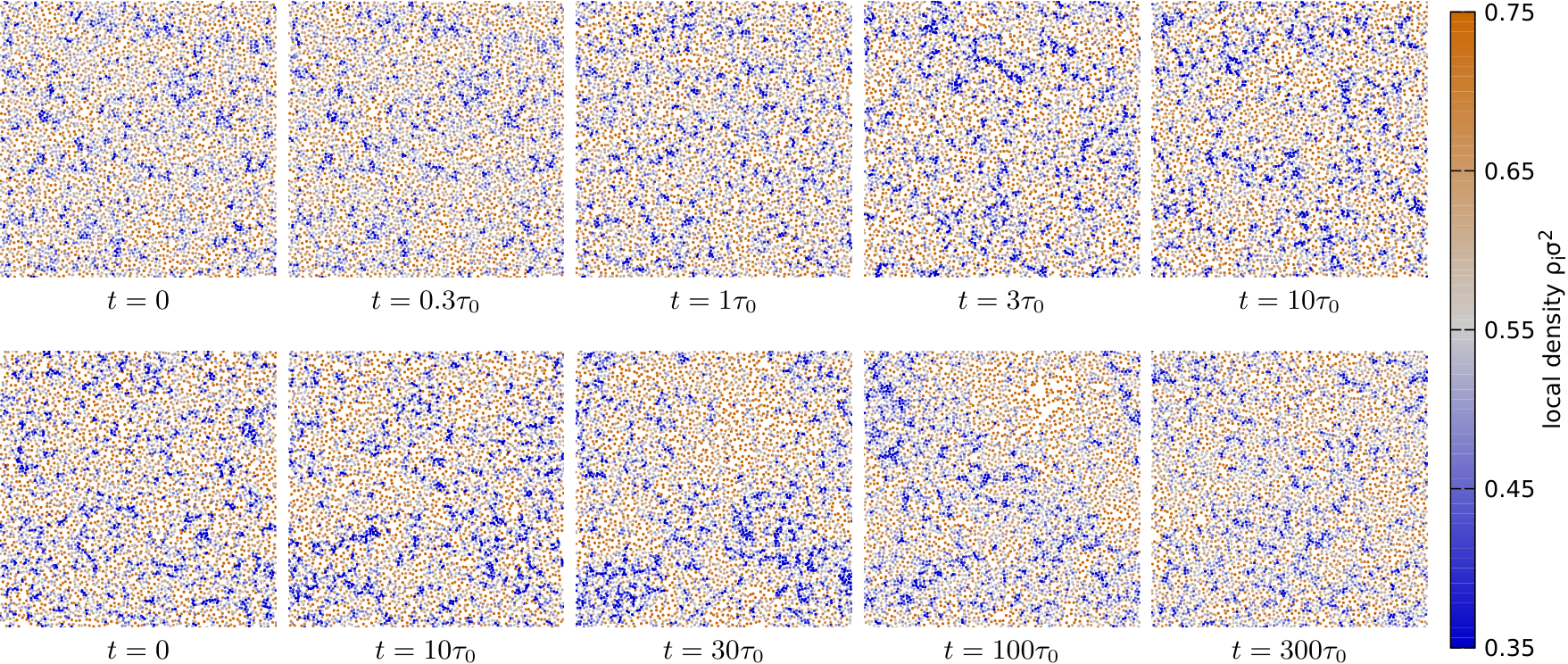}
		\caption{\label{fig:snap-bq}%
			Time series of top view simulation snapshots.
			Top row: Strong heating from \fromto13 (type~I).
			Bottom row: Strong cooling from \fromto31 (type~II).
			Amplitude change takes place at $t=0$. The color of each particle $i$ encodes its local density $\rho_i$ as depicted in the color bar.
			We calculate $\rho_i$ as the inverse area of its \ac{2d} Voronoi cell~\cite{sack1999handbook},
			which is the set of points closer to particle $i$ than to any other particle.
		}
	\end{figure*}
	
	The different types can even be distinguished when comparing simulation snapshots by eye.
	\figur{fig:snap-bq} shows two different time series, where the particle color encodes the local density.
	In the type~I simulation (top row) the system remains homogeneous and only the distribution of local densities of the particles becomes more heterogeneous with increasing $T$, which can be seen by the number of small dense patches increasing.
	This is a consequence of faster particles exploring the upper half of the box and hence displaying larger overlaps in projection (\cf Appendix).
	The bottom row shows the reverse process of strong cooling and has switched initial and final states.
	At intermediate times, however, we observe the formation of dense and dilute domains on the scale of the box size. (See, \eg the diluted region in the upper right part in the snapshots of $t=30\tau_0$ and $100\tau_0$.)
	This constitutes a clustering instability (see, \eg Refs.~\cite{goldhirsch1993clustering,poschel2005transient}) which is ultimately dissolved by the weak shaking at $A\fin$.
	The clustering instability is not necessarily but commonly observed in free cooling states.
	It therefore serves as an indicator of free cooling, and nicely illustrates the asymmetry between cooling and heating discussed before.

	\subsection{Large-scale correlations}\label{sec:quench-bulk-corr}
	
	\begin{figure*}
		\includegraphics[width=\thirdnl]{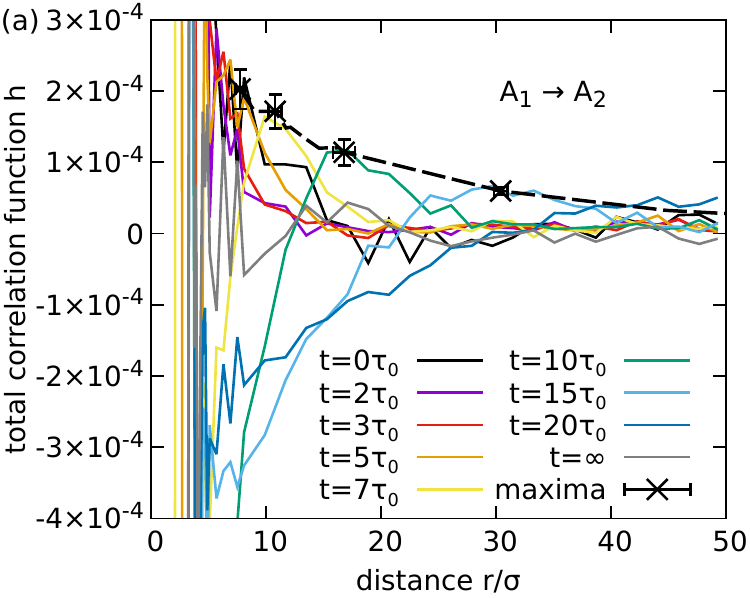}
		\includegraphics[width=\thirdnl]{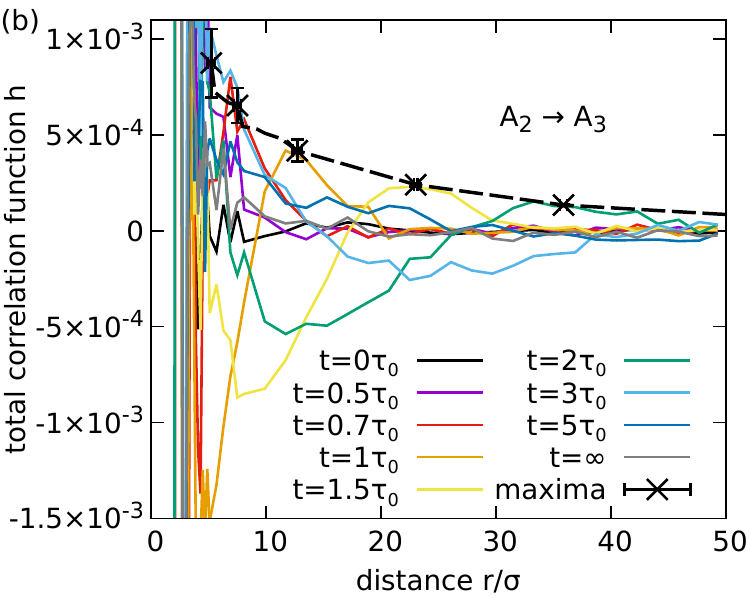}
		\includegraphics[width=\thirdnl]{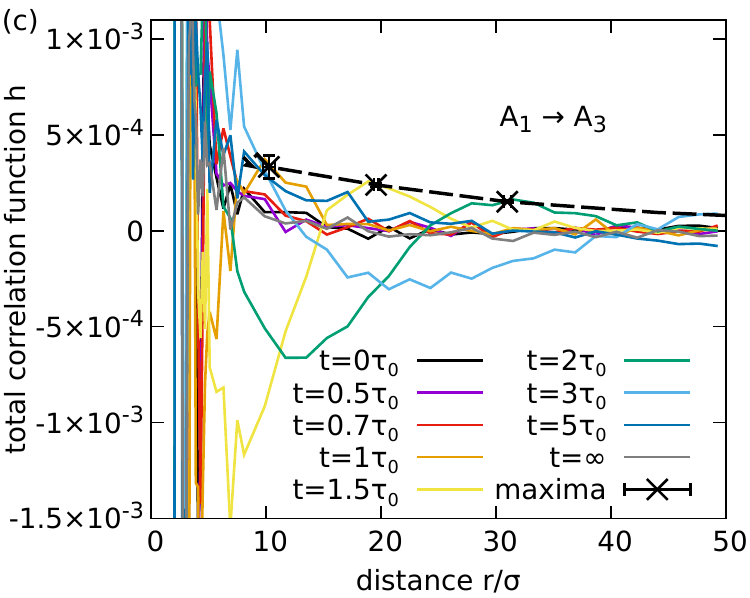}\\
		\includegraphics[width=\thirdnl]{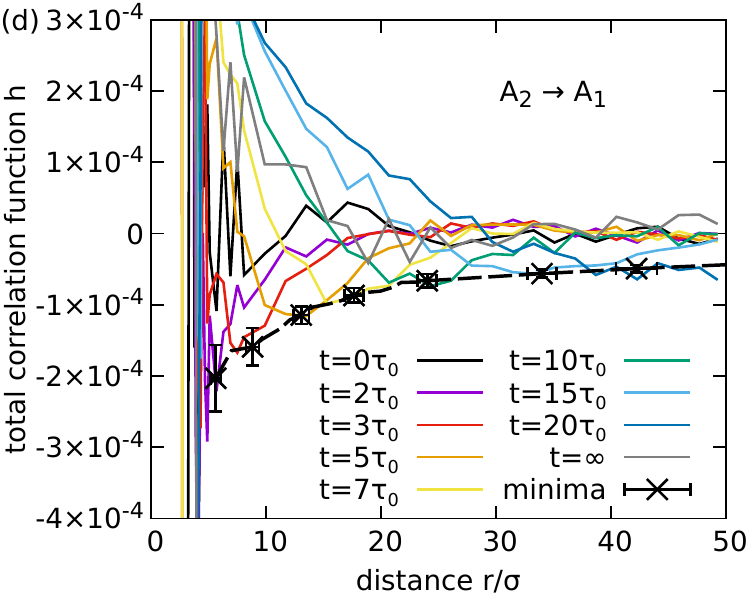}
		\includegraphics[width=\thirdnl]{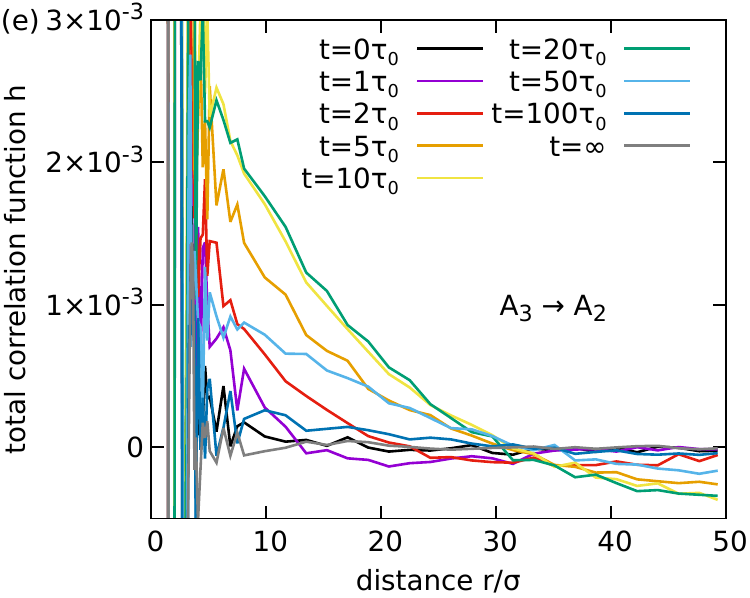}
		\includegraphics[width=\thirdnl]{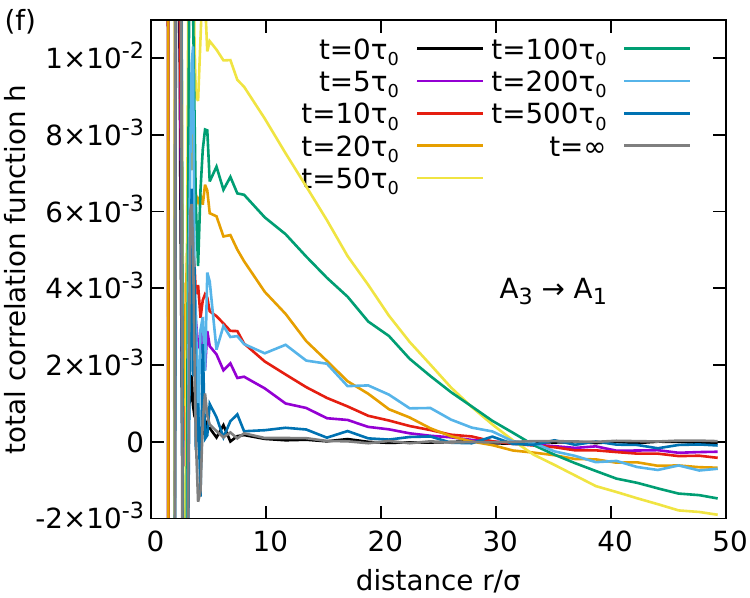}
		\caption{\label{fig:corr-bqA}%
			Total correlation function $h$ as a function of distance $r$, obtained with box size $L=100\sigma$ at a selection of times $t$ as indicated.
			A small value of $5\times 10^{-5}=0.25/N$ has been added to the data to correct the usual ${\cal O}(1/N)$ finite size effect of two point correlation functions~\cite{HMcD} such that $h(L/2,0)=0$.
			Top row panels \sub{a}--\sub{c} show heating protocols, bottom row panels \sub{d}--\sub{f} show cooling protocols.
			The left column \sub{a} and \sub{d} shows quenches between $A_1$ and $A_2$ (moderate), the middle column \sub{b} and \sub{e} between $A_2$ and $A_3$ (strong), and the right column \sub{c} and \sub{f} between $A_1$ and $A_3$ (strong).
			Black crosses in type~I quenches \sub{a}--\sub{d} mark the rightmost local extrema $(r\ex,h\ex)$ at the shown $t$, extracted from quadratic polynomial fits of $h(r)$ to that region.
			The correlation length $r\ex$ and the magnitude $h\ex\equiv h(r\ex)$ of correlations extracted in this manner are utilized in our subsequent analysis.
			Dashed black lines connect fitted extrema of all $t$ \bracket{also of $t$ for which the $h(r)$ are not shown}.
		}
	\end{figure*}
	
	Now we turn to two point correlations measured by the transient
	total correlation function~\cite{HMcD}
	\begin{equation}
		h(r,t) \equiv\left\langle \frac{1}{\rho^2}\sum_{i=1}^{N}\sum_{j\neq i}^{N}\delta(\vec r_i - \vec r)\delta(\vec r_j - \vec r')\right\rangle-1\,,
	\end{equation}
	where $\delta$ is the Dirac delta distribution, $\vec r$ and $\vec r'$ are two \ac{2d} position vectors, and $r\equiv |\vec r-\vec r'|$.
	\figur{fig:corr-bqA} shows time series of $h$ as a function of $r$ for all considered quench protocols.
	In the range $r<5\sigma$ there are exponentially decaying oscillations (off scale),
	which constitute the fluid structure also present in the steady state \bracket{\cf \subfig{fig:bs}{b}}.
	The focus of this work is not this well-known feature of any dense liquid~\cite{HMcD} but rather the transient contributions that are observed at larger distances $r>5\sigma$.
	
	In the initial state $t=0$ (black curves) there are no large-scale correlations.
	After the quench, however, these build up in time and decay to zero again as $t\rightarrow\infty$.
	There are two rather distinct types of behaviors for type~I and II protocols as classified in Sec.~\ref{sec:quench-bulk-global}.
	In type~I simulations \bracket{\subfig{fig:corr-bqA}{a}--\sub{d}} we find oscillating correlations with local maxima and/or minima.
	(The monotony is discussed in Sec.~\ref{sec:quench-bulk-comp})
	The extrema are propagating in time towards $r\rightarrow\infty$ and are of the order of at most $\sim 10^{-3}$ in magnitude.
	The shape of each function does not exhibit finite-size scaling, \ie it is stable against utilizing different box sizes (aside from boundary effects at $r \approx L/2$).
	Type~II protocols \bracket{\subfig{fig:corr-bqA}{e} and \sub{f}} exhibit larger correlations up to $10^{-2}$ in magnitude that are not propagating and are positive in sign.
	The fact that the correlations do not vanish for $r\rightarrow L/2$, but approach a small negative value, indicates density inhomogeneities at the scale of the box size, and clearly constitutes a finite-size effect.
	This confirms the conclusions from the inspection of the snapshots and characterization of type~II behavior as inelastic collapse.

	\begin{figure}
		\includegraphics[width=\halfnl]{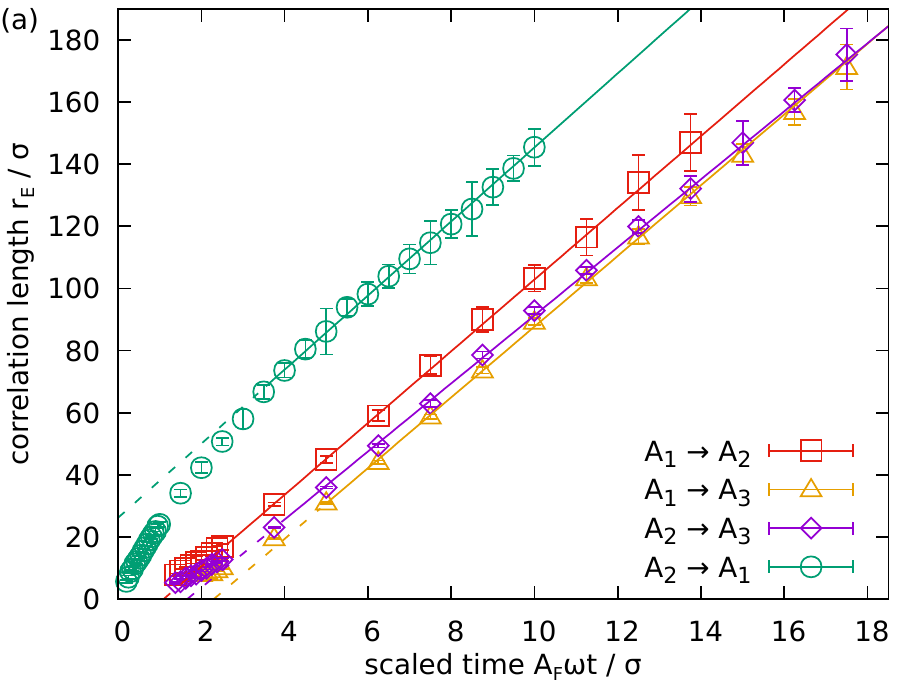}\\
		\includegraphics[width=\halfnl]{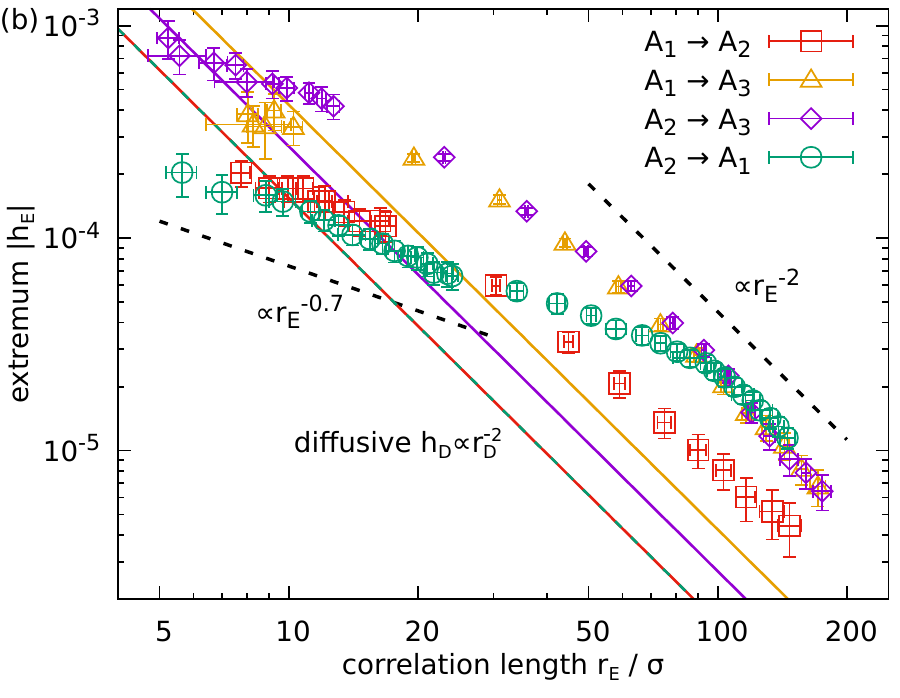}
		\caption{\label{fig:XS-CS-t}%
			\sub{a} Correlation length $r\ex$ of the total correlation function $h$ as a function of time $t$ (multiplied by the post-quench peak velocity of the oscillating plates $A\fin\omega$ for comparability), extracted from type~I protocols \bracket{\subfigs{fig:corr-bqA}{a}--\sub{d}}.
			Straight lines are linear fits obtained from the data ranges, where the lines are continuous.
			\sub{b} Absolute values $|h\ex|$ of the magnitude of correlations
			as a function of $r\ex$ on double logarithmic scales \bracket{same data points as black crosses and dashed lines in \subfigs{fig:corr-bqA}{a}--\sub{d}}.
			Straight lines are the theoretical predictions $h\theo(r\theo)$ for diffusive systems with instantaneous temperature quenches according to \Eq{eq:CS-XS}.
			The black dashed lines corresponding to power laws with exponents $-0.7$ and $-2$ are guides to the eye.
			Data points of both panels at $r<50\sigma$ have been obtained in simulation boxes with $L=100\sigma$, data points at $r>50\sigma$ have been obtained in simulation boxes with $L=400\sigma$.
		}
	\end{figure}
	
	The remainder of the section is devoted to quantitative evaluation of the type~I behavior
	via the length and magnitude of the correlations.
		The correlation length $r\ex$ is measured in terms of the positions of the rightmost local extrema \bracket{marked with black crosses in \subfigs{fig:corr-bqA}{a}--\sub{d}}.
	\subfigure{fig:XS-CS-t}{a} shows the $r\ex$ as functions of $t$.
	We identify two distinct regimes, namely a short-time regime during which $T$ still adjusts to $A\fin$ (\cf \fig{fig:PT-bqA}), and a long-time regime at constant $T$.
	In the short-time regime, the propagation velocity $v\ex$ of the extremum increases or decreases as $T$ increases or decreases in heating or cooling protocols (\cf \fig{fig:PT-bqA}).
	As $T$ relaxes to the final steady-state value, $v\ex$ takes a constant value approximately proportional to $A\fin$, which lies in the range
	\begin{equation}
		v\ex=(11.4\pm 0.5)A\fin\omega
	\end{equation}
	for the different protocols.
	Hence we obtain the dynamic scaling exponent $\alpha=1$ (defined via $r\ex\propto t^\alpha$) of ballistic motion.
	
	A physical interpretation of $v\ex$ could be provided by the following possible origin of the correlations,
	which--in the case of heating--is similar to bursts caused by collisions of heavy particles reported in Ref.~\cite{rivas2012characterization}.
	Upon heating (weak cooling), particles are accelerated (decelerated) in the $z$ direction by the oscillating plates to a velocity $\propto A\fin\omega$.
	Accelerated (decelerated) particles may transfer their kinetic energy to (recover vertical kinetic energy from) horizontal directions in particle-particle collisions,
	which creates pairs of excess (depleted) momenta in opposite directions of the involved particles.
	These pairs of momenta induce particle currents, which create correlations that are transported
	through the granular medium.
	In this manner, momentum is transferred without loss to other particles, as momentum is conserved in collisions.
	The propagation speed in the long-time limit (\ie at saturated temperature) can be estimated as the mean velocity $v_T\equiv\sqrt{{T\fin}/{m_\mathrm{p}}}$ of particles, plus a contribution of the distance covered while momentum is passed on in collisions,
	\begin{equation}
		v\ex\approx v_T+s r_\mathrm{c}\,.
	\end{equation}
	Here $s$ is the mean projected distance of colliding particles and $r_\mathrm{c}$ is the rate of particle-particle collisions.
	In steady-state simulations we measured values in the ranges of $v_T$ from $2.8A\fin\omega$ to $3.0A\fin\omega$ and of $r_\mathrm{c}$ from $9.3A\fin\omega/\sigma$ to $10.7A\fin\omega/\sigma$ for the considered amplitudes.
	$s$ has not been sampled explicitly but can be calculated from the other quantities to be in the range from $0.82\sigma$ to $0.93\sigma$, which is reasonable in this quasi-\ac{2d} setup.
	In the intermediate regime the velocity may be reduced (enhanced) due to the reduced (enhanced) temperature.
	This mechanism requires that particles are thermalized at a variety of rates, which is not given in the case of strong cooling, where $T\fin\ll T\init$, which means that plates are practically immovable and energy loss is dominated by particle inelasticity.
	
	We stress that $v\ex$ is not the speed of sound $c_\mathrm{s}$ of the final steady-state fluid, which we calculated in supplemental steady-state simulations via the dynamic structure factor according to the method described in Ref.~\cite{HMcD}
	as $c_\mathrm{s}(A_1)=(7.8\pm0.4)A\fin\omega$ and $c_\mathrm{s}(A_3)=(4.2\pm0.2)A\fin\omega$.
	The fact that $c_\mathrm{s}\not\propto A\fin\omega$ clearly disqualifies the speed of sound as possible interpretation for $v\ex$.
	
	The magnitude $h\ex$ of correlations, measured as the value of the correlation function at the extremum \bracket{see \subfig{fig:XS-CS-t}{b}}, also exhibits a crossover like $r\ex$ with the same crossover times.
	In each of the regimes the dynamics is describable by an algebraic scaling.
	In the short-time regime $h\ex\propto r\ex^{-0.7}$
	and in the long-time regime
	$h\ex\propto r\ex^{-2}$
	for all protocols,
	which yields the scaling exponent of the correlation strength $\beta=-2$.

	\subsection{Comparison to Brownian dynamics}\label{sec:quench-bulk-comp}
	
	We now compare our results to diffusive systems for which a theory has recently been  developed~\cite{rohwer2017transient,rohwer2018nonequilibrium} to describe fluctuation induced correlations after instantaneous temperature changes.
	We stress that we do not apply this theory to the present simulation results expecting quantitative agreement, but rather compare to the analytical results for Brownian systems to detect universalities and system-specific properties.
	The theory predicts a Gaussian distribution of the correlations,
	\begin{equation}
		\label{eq:C-X-t-anal}
		h(r,t)=\frac{S\init-S\fin}{\rho}\frac{\exp\left({-\frac{r^2}{2r\theo(t)^2}}\right)}{\sqrt{2\pi r\theo(t)^2}^{\,d}}\,,
	\end{equation}
	where $S\init$ and $S\fin$ are the zero wavelength limits of the static structure factors of the initial and final steady-state fluid, respectively, and $r\theo$ is the correlation length.
	For quantitative comparison, we extract the inflection point of this function, giving us the typical strength and length scale of correlations.
	The typical strength is obtained by setting $r=r\theo$,
	\begin{equation}
		\label{eq:CS-XS}h\theo(r\theo)=\frac{S\init-S\fin}{\rho\sqrt{e}2\pi r\theo^2}\propto r\theo^{-2}\,.
	\end{equation}
	yielding a scaling exponent $\beta=-2$.
	Note that this is independent of the dynamical scaling of $r\theo$ and only depends on the spatial dimension.
	Therefore, $\beta$ can be seen as a geometric property
	that ensures constant normalization of the Gaussian distribution.
	The correlation length as a function of time reads
	\begin{equation}
		\label{eq:XS-t}r\theo(t)=\sqrt{4D\fin t}\propto t^{1/2}\,,
	\end{equation}
	with the long-time single-particle diffusion coefficient $D\fin$ of the final steady state.
	This implies $\alpha=1/2$--the dynamic scaling exponent of diffusive motion.
	
	The prefactor in \Eq{eq:CS-XS} is proportional to the difference of the initial and final steady-state static structure factors, which are connected to the respective compressibilities $\chi$ via
	$S=\rho T \chi\equiv T \partial \rho/\partial P$.
	Hence, we only expect a nonzero effect for thermal particles with variable softness.
	In the present setup we employ hard spheres, which are athermal by themselves.
	As scrutinized in the Appendix, however, we create an effective softness via variation of the stratification of the particles at different $A$, which influences their overlaps in the \ac{2d} projection.
	This makes it possible to observe the predicted effect in our system despite the hard core model.
	Note that we use the term ``softness'' for structural properties of the fluid rather than material properties of the individual particles.
	Overlaps due to actual compression of glass or metal spheres in experiments or in different simulation models would be orders of magnitude smaller than the overlaps due to stratification.
	
	A first notable observation upon comparing our results to the diffusive case, is the rather different functional form of $h$.
	While the Brownian theory predicts a universal Gaussian shape of the correlations, we observed a more complicated function shape with oscillating behavior that depends on the applied protocol (see also below).
	Therefore, we only compare the scaling exponents of the extracted extrema shown in \fig{fig:XS-CS-t} to \Eqs{eq:CS-XS} and \eqref{eq:XS-t}, and not the prefactors.
	
	Even though the theory is not directly applicable here, the value $\alpha=1$ obtained in our simulations is remarkable.
	$T$ is fully relaxed in the long-time regime and thus one could expect the post-quench fluid to behave like a steady-state fluid.
	Indeed, we performed preparatory steady-state simulations
	where we find diffusive motion of the individual particles for $r>50\sigma$.
	However, the correlation function exhibits the ballistic scaling of propagating waves.
	This confirms the interpretation as a collective phenomenon, in which information is not carried by the individual particles, but transferred between particles in collisions, as proposed in Sec.~\ref{sec:quench-bulk-corr}.
	
	The picture is quite different when considering $\beta$.
	Here the long-time limit $\beta=-2$ fits the diffusive theory well.
	This universality supports the notion that $\beta$ is a purely geometric quantity that is independent of the dynamic details and designates the correlations as a conserved quantity in steady state.
	The scaling $|h\ex|\propto r\ex^{-0.7}$ during thermalization, on the other hand, indicates a normalization which increases in time or in other words a source of correlations.
	This observation supports the picture of a build-up of correlations during temperature equilibration as proposed before.

	\subsection{Universal functional shape}\label{sec:quench-bulk-shape}
	
	\begin{figure}
		\includegraphics[width=\quarternl]{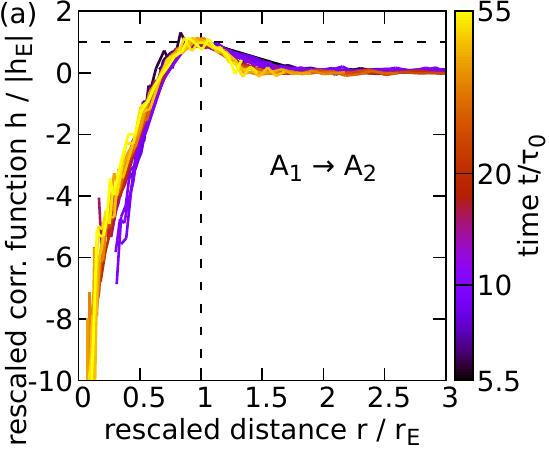}
		\includegraphics[width=\quarternl]{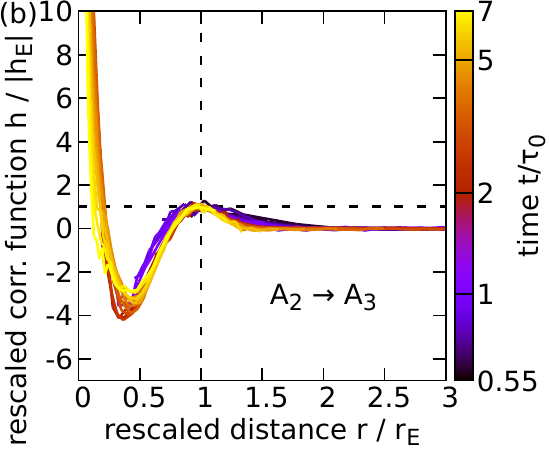}\\
		\includegraphics[width=\quarternl]{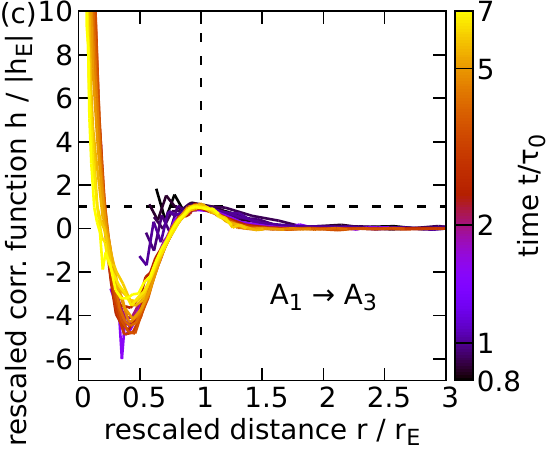}
		\includegraphics[width=\quarternl]{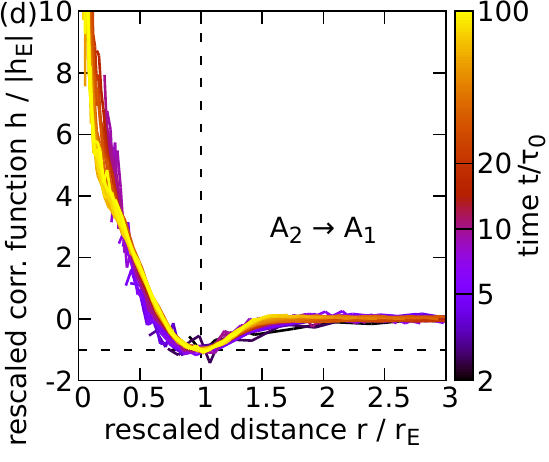}
		\caption{\label{fig:C-X-rescaled}%
			The same data as \subfigs{fig:corr-bqA}{a}--\sub{d}, but with abscissa and ordinate rescaled by correlation length $r\ex(t)$ and correlation magnitude $|h\ex(t)|$, respectively \bracket{such that the extrema collapse at $(1,1)$ or $(1,-1)$}.
			Protocols are indicated in the labels.
			Only data exhibiting an extremum in $h$ is shown.
			Panels \sub{a} and \sub{b} show moderate heating and cooling, respectively.
			Panels \sub{c} and \sub{d} show strong heating.
			Time is encoded in the color of each curve as depicted in the colored bars of the panels.
		}
	\end{figure}
	
	In order to characterize its features, we collapse $h(r)$ for different $t$ by rescaling abscissa and ordinate with the values of the extrema, \ie we plot $h / |h\ex|$ versus $r / r\ex$ as shown in \subfigs{fig:C-X-rescaled}{a}--\sub{d} for the different quench protocols.
	By definition, this scaling function has a local maximum at $(1,1)$ for heating or minimum at $(1,-1)$ for cooling protocols.
	What is striking here is the stability of the function shapes, which is in sharp contrast to the strong crossover of the scaling exponents.
	There are only slight shifts of the functions left and right of the main extrema at the times of the crossover.
	
	At moderate heating \bracket{\subfig{fig:C-X-rescaled}{a}} only a single maximum is present.
	For $r\rightarrow 0$, $h$ attains negative values and for $r\rightarrow\infty$, $h$ decays towards zero.
	
	The strong heating protocols \sub{b} and \sub{c} both show the same qualitative behavior,
	which differs from moderate heating by an additional local minimum at $\approx(0.4,-4)$ before or $\approx(0.5,-3)$ after the crossover.
	Consequently there is an additional zero at $r/r\ex\approx0.2$ and $h>0$ for $r\rightarrow0$.
	This new feature could be attributed to an additional process that only takes place when the fluid temperature is changed strongly.
	The feature is located at smaller $r$ than the original extremum, indicating that the additional process takes place at a later time than the process 
	that creates the maximum at $(1,1)$.
	The data, however, does not reveal the nature of this process.
	
	Moderate cooling \bracket{\subfig{fig:C-X-rescaled}{d}} exhibits exactly the same behavior as moderate heating but with the opposite sign.
	This is a clear signature for a linear response regime.

	\section{Casimir geometry}\label{sec:casimir}
	
	\begin{figure*}
		\includegraphics[width=\thirdnl]{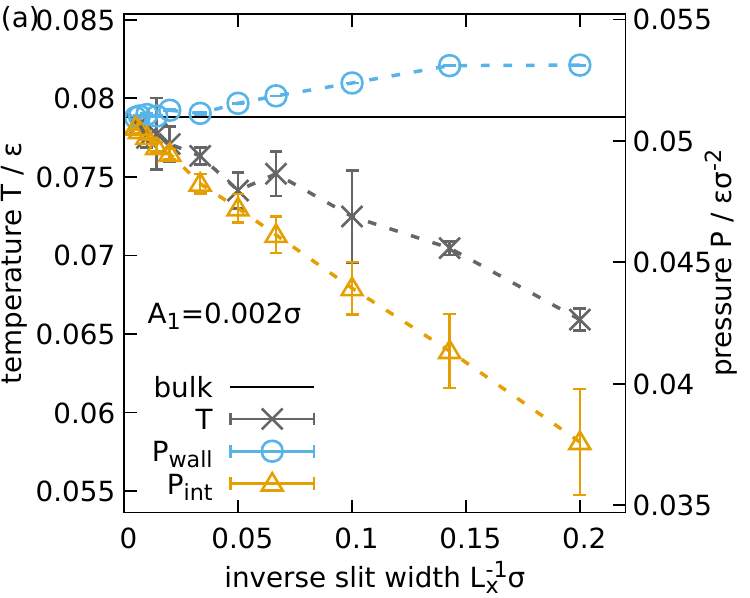}
		\includegraphics[width=\thirdnl]{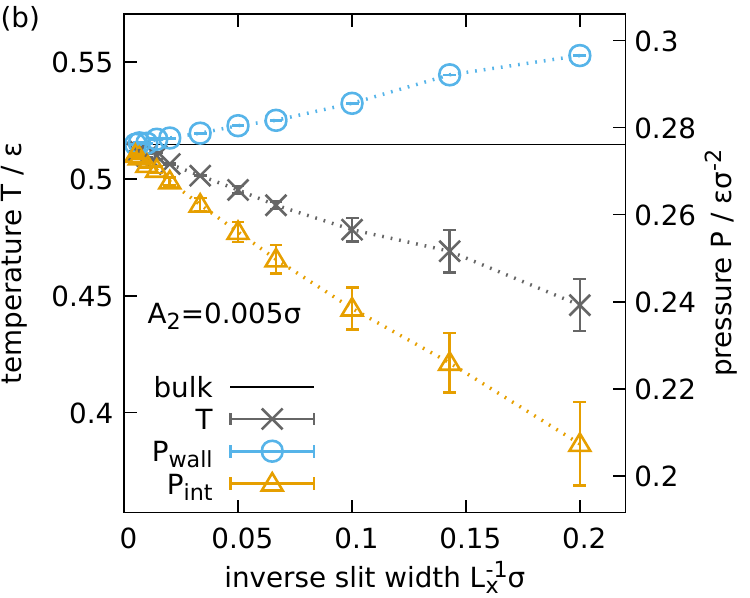}
		\includegraphics[width=\thirdnl]{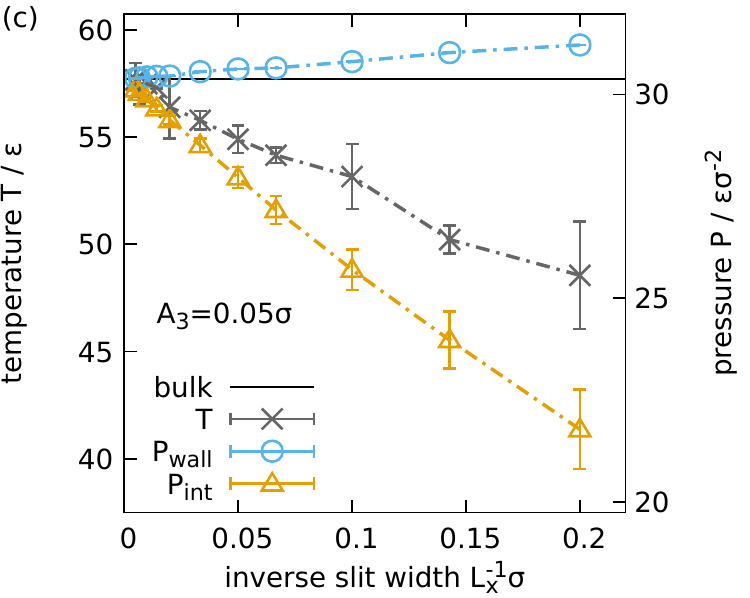}\\
		\caption{\label{fig:P-cs}%
			Temperature $T$, internal pressure \pint, and side wall pressure \pwall in the slit geometry as functions of inverse slit width $L_x^{-1}$ at steady state.
			\sub{a} Amplitude $A_1$, \sub{b} $A_2$, and \sub{c} $A_3$.
			The temperature scale (left ordinate) and pressure scale (right ordinate) are adjusted such that the origins (off scale) and bulk values of $T$ and \pint (marked by straight black lines) coincide.
			Lines connecting data points are guides to the eye.
		}
	\end{figure*}
	
	This section treats the setup with side walls as depicted in \subfig{fig:set}{b}.
	The aim here is to report boundary effects on global observables and to determine whether these are caused by the bulk post-quench correlations described in the previous section as is the case in diffusive systems~\cite{rohwer2017transient}.
	
	We start by describing confinement effects in the steady state by means of $T$, \pint and the pressure on the side walls,
	\begin{equation}
		\pwall\equiv\left\langle\frac{1}{2L_y\Delta t}\sum_{\mathrm{PW}}\Delta\vec{p}_i\cdot\vec{n}_{\mathrm{w}}\right\rangle\,,
	\end{equation}
	where the sum is performed over all particle-wall collisions of any particle $i$ with either of the side walls (with normals $\vec{n}_{\mathrm{w}}=\pm\vec{e}_x$) during $\Delta t$.
	\figur{fig:P-cs} shows $T$, \pint, and \pwall as functions of $L_x^{-1}$.
	We observe a linear dependence on $L_x^{-1}$ in all three functions.
	Extrapolations to $L_x\rightarrow\infty$ agree with the values $T_\mathrm{bulk}$ and $P_\mathrm{bulk}$ of temperature and internal pressure, respectively, of separate bulk simulations performed beforehand (dashed lines).
	Linear fitting and averaging over amplitudes yields
	\begin{equation}
		\label{eq:sfy-cs}
		\begin{split}
			T(L_x)=&T_\mathrm{bulk}[1-(0.8\pm 0.1)\sigma L_x^{-1}]\,,\\
			\pint(L_x)=&P_\mathrm{bulk}[1-(1.4\pm 0.1)\sigma L_x^{-1}]\,,\\
			\pwall(L_x)=&P_\mathrm{bulk}[1+(0.3\pm 0.1)\sigma L_x^{-1}]\,.
		\end{split}
	\end{equation}
	\bracket{Note that $\pint(L_x)$ and $\pwall(L_x)$ converge to the same value in the limit $L_x\rightarrow\infty$.}
	$T$ decreases with decreasing $L_x$ and \pint follows the temperature behavior.
	\pwall, however, increases with decreasing $L_x$, which constitutes a nontrivial effect:
	In a setup, where the side walls were placed in an infinite system with fluid both inside and outside the side walls, the exterior domain would exert a side wall pressure of the extrapolated value.
	Hence, there would be a net \textit{expanding} force on the side walls even though the granular temperature is smaller at the interior.
	Qualitatively, the finite-size scaling can be described in terms of an excess particle accumulation at the side walls~\cite{rohwer2018nonequilibrium}
	that changes the density between the walls by a contribution $\propto L_x^{-1}$.
	
	\begin{figure}
		\includegraphics[width=\quarternl]{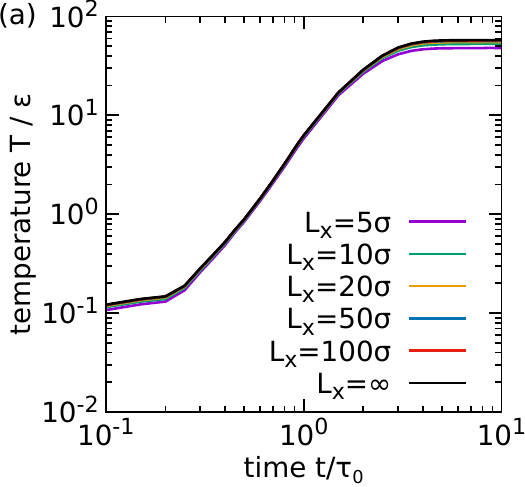}
		\includegraphics[width=\quarternl]{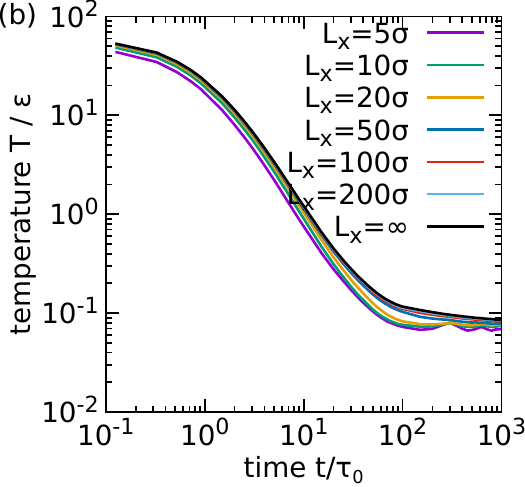}\\
		\includegraphics[width=\quarternl]{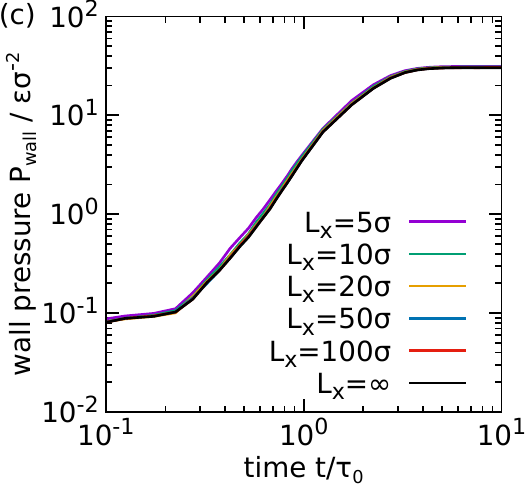}
		\includegraphics[width=\quarternl]{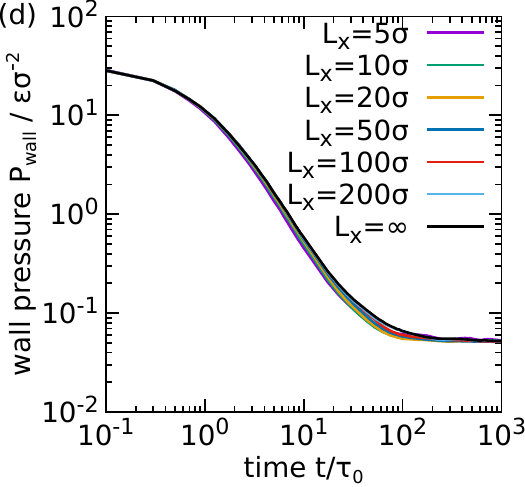}\\
		\includegraphics[width=\quarternl]{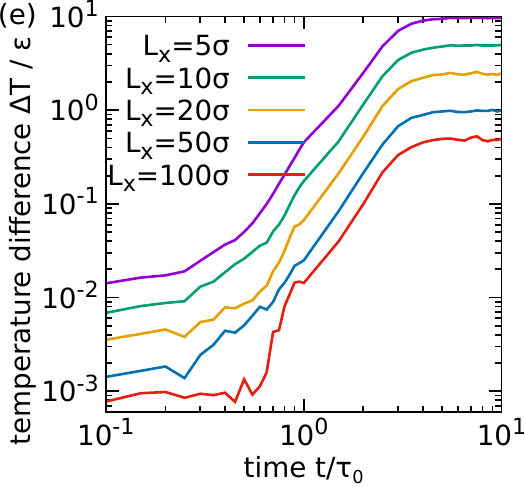}
		\includegraphics[width=\quarternl]{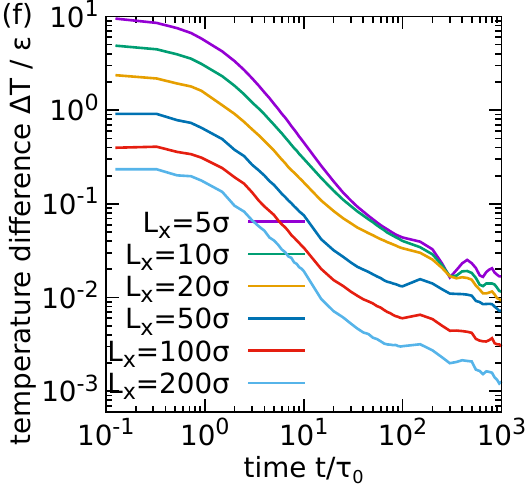}\\
		\includegraphics[width=\quarternl]{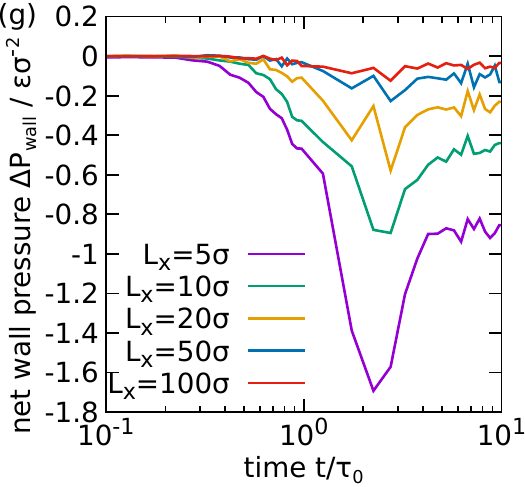}
		\includegraphics[width=\quarternl]{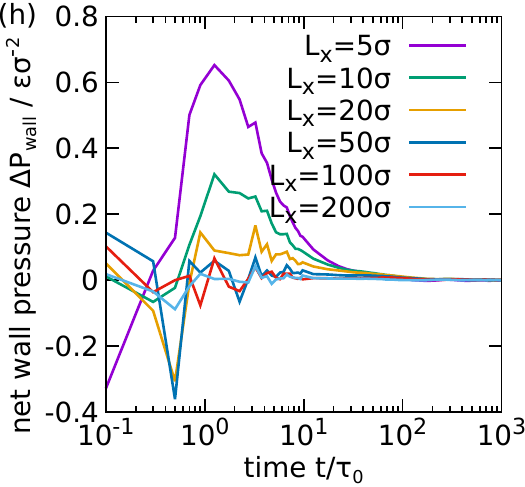}\\
		\caption{\label{fig:P-cq}%
			Data for temperature $T$ and side wall pressure $\pwall$ as functions of time $t$ after the quench (logarithmic scale) for several slit widths $L_x$ as indicated in each panel.
			\sub{a} and \sub{b} $T$ on logarithmic scale.
			\sub{c} and \sub{d} $\pwall$ on logarithmic scale.
			\sub{e} and \sub{f} Temperature difference $\Delta T\equiv T(\infty)-T(L_x)$ between the exterior and interior of the slit on logarithmic scale.
			\sub{g} and \sub{h} Net side wall pressure $\Delta \pwall\equiv \pwall(\infty)-\pwall(L_x)$ (\ie difference between pressures on outer and inner surfaces of the side walls).
			Left column  \sub{a}, \sub{c}, \sub{e}, and \sub{g}: Strong heating \fromto13 (type~I).
			Right column \sub{b}, \sub{d}, \sub{f}, and \sub{h}: Strong cooling \fromto31 (type~II).
			Data for $L_x=\infty$ is extrapolated from finite $L_x$ at each time, as was done for the steady-state values (\cf \fig{fig:P-cs}).
		}
	\end{figure}
	
	Next, we turn to quenches of the Casimir geometry, and investigate whether there is an effect beyond these steady-state confinement effects.
	The dynamics of $T$ is shown for a type~I quench in \subfig{fig:P-cq}{a} and a type~II quench in \subfig{fig:P-cq}{b}.
	As in the bulk setup (cf.\ \fig{fig:PT-bqA}), $T$ relaxes to the steady-state value $T\fin$ exponentially in the type~I quench but algebraically in the type~II quench.
	Other type~I quenches between $A_1\leftrightarrow A_2$ (not shown) also show the same behavior as \subfig{fig:P-cq}{a}.
	\pwall shown for heating in \subfig{fig:P-cq}{c} and cooling in \subfig{fig:P-cq}{d} exhibits similar behavior as $T$.
	The main difference is the inverted finite-size scaling (\ie deviation from the limit $L_x\rightarrow\infty$) of the initial and final states, which is consistent with the steady-state results (\cf \fig{fig:P-cs}).
	
	The temperature differences $\Delta T$ between the exterior and interior \bracket{\subfig{fig:P-cq}{e} and \sub{f}} transition monotonically from the initial to the final steady-state values plotted in \subfig{fig:P-cs}{a} and \sub{c}.
	The net side wall pressure $\Delta \pwall$ \bracket{\subfig{fig:P-cq}{g} and \sub{h}}, however, shows an undershoot or overshoot in the case of heating or cooling, respectively.
	The undershoots in the heating simulations \bracket{\subfig{fig:P-cq}{g}} for all $L_x$ take place at $t\approx2-3\tau_0$, which is roughly the time at which the fluid is fully heated \bracket{\cf \subfig{fig:P-cq}{a}}.
	The times $t\approx1-2\tau_0$ of the overshoots at cooling \bracket{\subfig{fig:P-cq}{h}} coincide with the start of the cooling of the fluid \bracket{\cf \subfig{fig:P-cq}{b}}.
	Moderate heating \fromto12 shows the same behavior as strong heating and moderate cooling \fromto21 shows the same behavior as strong cooling (not shown).
	The extremum is, however, not as pronounced upon moderate amplitude changes.
	We would like to stress that the qualitative behavior here does not correspond to the type-I and -II classification of the previous section but depends on whether the fluid is cooled or heated.

	\begin{figure*}
		\includegraphics[width=\thirdnl]{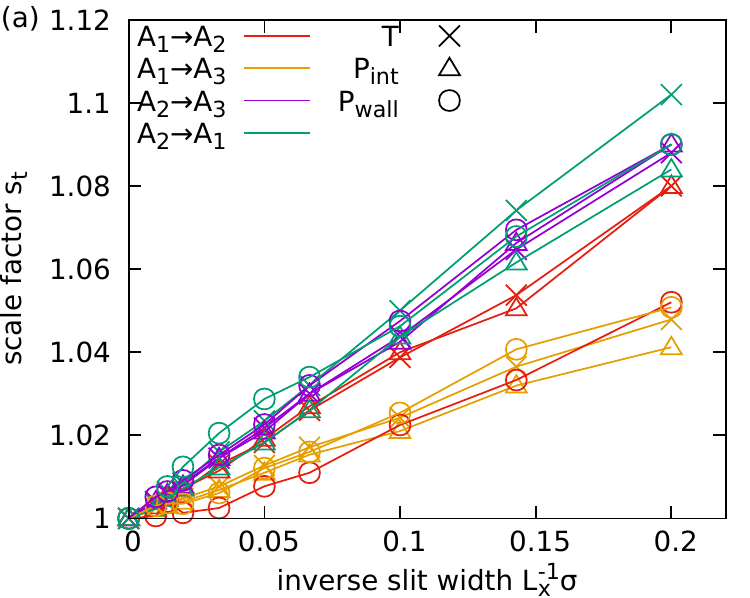}
		\includegraphics[width=\thirdnl]{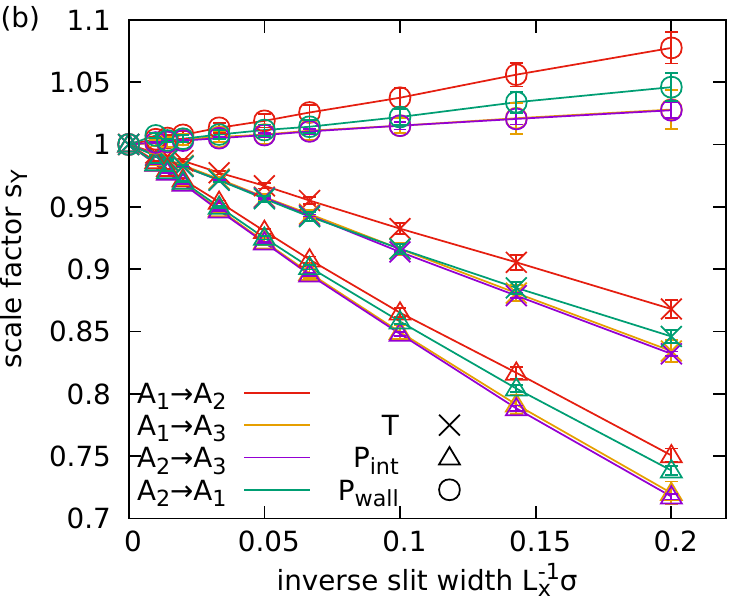}
		\includegraphics[width=\thirdnl]{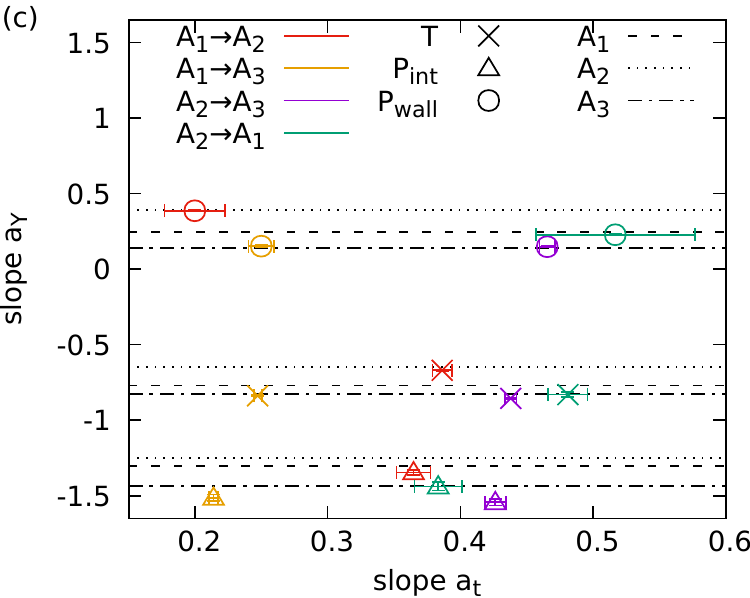}\\
		\caption{\label{fig:scaling-cq}%
			\sub{a} Scale factor \sft of time as a function of inverse slit width $L_x^{-1}$.
			Errors (omitted for clarity) are of magnitude $\sigma_{\sft}=0.05L_x^{-1}\sigma$\,--\,$0.2L_x^{-1}\sigma$.
			\sub{b} Same as \sub{a}, but for scale factors \sfo  of the observables $Y=T,\pint,\pwall$.
			Data points for $\fromto13$ (yellow) are hardly visible as they mostly coincide with data points for $\fromto23$ (purple).
			Lines in \sub{a} and \sub{b} are guides to the eye.
			Errors $\sigma_{\sft}$ and $\sigma_{\sfo}$ are obtained via the criteria $E(\sft,\sfo\pm\sigma_{\sfo})\equiv 2E(\sft,\sfo)$ and $E(\sft\pm\sigma_{\sft},\sfo)\equiv 2E(\sft,\sfo)$.
			\sub{c} Slopes $a_Y$ of linear fits to \sfo as functions of the slopes $a_t$ of linear fits to \sft.
			Lines show the steady-state values of $a_Y$ \bracket{\cf \Eq{eq:sfy-cs}} at $A_1$ (dashed), $A_2$ (dotted), and $A_3$ (dashed-dotted).
			The legends show the encoding of the respective collapsed observables via symbol types, and the quench protocols via colors.
		}
	\end{figure*}
	
	In general, nonmonotonic behavior of the pressure and temperature may indicate a separation of timescales as regards dynamics in the vicinity of the surfaces and dynamics of the bulk. Overshoots of this type have, \eg been observed in drift-diffusion systems~\cite{khalilian2020interplay}.
	To explain the net pressure overshoot in our system, we collapse the data and obtain the finite-size scaling.
	We assume that the data follows the scaling relation
	\begin{equation}
		\label{eq:scaling-walls}
		Y(t/\sft(L),L)=\sfo(L)Y(t,\infty)
	\end{equation}
	for all considered observables $Y=T, \pint, \pwall$,
	with scaling factors \sft and \sfo describing the scaling of time and of the final steady-state value, respectively.
	The scaling factors are determined by minimizing an error function defined via
	\begin{equation}
		E(\sft,\sfo) \equiv \frac{1}{t\fin}\int_{0}^{t\fin} \mathrm{d}t\left[\frac{Y(t/\sft,L)}{\sfo Y(t,\infty)} -1\right]^2,
	\end{equation}
	with an upper bound of integration $t\fin$, which is in practice given by the range of simulation data.
	This error function is defined such that $E=0$ if \Eq{eq:scaling-walls} is satisfied and $E>0$ otherwise.
	With the input simulation data for $Y(t,L)$ and $Y(t,\infty)$ being noisy,
	the true values of \sft and \sfo are the ones that minimize $E$.
	These are determined simply by sampling a fine grid in the $\sft \sfo$ plane and taking the minimum value.
	The position of the minimum $(\sft,\sfo)$ does not depend on the choice of $t\fin$ as long as $t\fin$ is greater than the time the fluid needs to relax.
	
	We judge the quality of this numerical data collapse by the minimal value of $E$ being small.
	This criterion is fulfilled for type~I protocols with $E < 0.001 (L_x/\sigma)^{-1}$.
	For type ~II protocols, however, we obtain values of up to $E \approx 0.1 (L_x/\sigma)^{-1}$.
	Therefore, we conclude that the finite-size scaling of cluster instability is not captured adequately by the simple two-parameter scaling of \Eq{eq:scaling-walls}.
	
	\subfigures{fig:scaling-cq}{a} and \sub{b} show \sft and \sfo as functions of $L_x^{-1}$.
	To a good approximation all scaling factors are linear functions of $L_x^{-1}$ with slopes $a_t$ and $a_Y$ plotted against each other in \subfig{fig:scaling-cq}{c}.
	The $a_Y$ are equal to our steady-state results in \Eq{eq:sfy-cs} within the uncertainties (which serves as an additional consistency check for the minimization procedure).
	
	Values for $a_t$ are in the range $0.2\leq a_t \leq 0.5$ for the different quench protocols, which implies an enhanced relaxation in the presence of side walls.
	With this, the nonmonotonic $\pwall$ can be explained as follows.
	The undershoot in the side wall pressure observed upon increasing $A$ \bracket{\subfig{fig:P-cq}{g}} takes place when the fluid between the side walls is already completely heated, while the pressure of the fluid outside is smaller as it is still adjusting to the post-quench amplitude.
	The overshoot in the reverse process \bracket{\subfig{fig:P-cq}{h}} takes place when the inner fluid has started to cool faster than the fluid outside.
	The obtained values for the slopes of the scaling factors also explain why the overshoot is not observed in $T$ or \pint:
	Here the overshoot through the enhanced relaxation is overlaid by decreasing of the (initial and final) steady-state values (as $a_t<0$ and $a_Y<0$ and $|a_Y|>|a_t|$ for $Y=T,\pint$).
	The slightly positive value of $a_{\pwall}$, on the other hand, slightly enhances the overshoot.

	\begin{figure}
		\begin{center}
			\includegraphics[width=\halfnl]{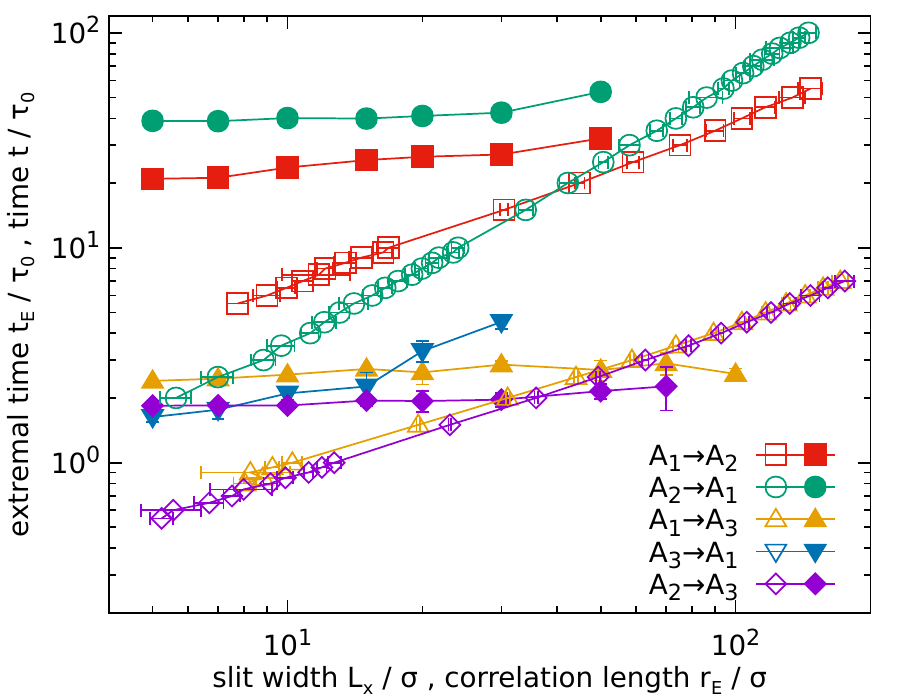}
		\end{center}
		\caption{\label{fig:t-XL}%
			Extremal times $t\ex$ of the side wall pressure under- or overshoots (filled symbols) versus slit width $L_x$ and time $t$ \bracket{open symbols; same data as \subfig{fig:XS-CS-t}{a}} versus correlation length $r\ex$ of the total correlation function $h$.
			Same quench protocols are depicted as same symbol shapes and colors.
			Lines are guides to the eye.
		}
	\end{figure}
	
	In order to affirm that the bulk correlations described in Sec.~\ref{sec:quench-bulk} are not the origin of the transient net pressure overshoot, we compare the dynamic scaling of both effects.
	To this end, we extract the times $t\ex$ at which the extrema of $\Delta \pwall$ are attained by fitting quadratic functions in the vicinities of the extrema (displayed in \fig{fig:t-XL}).
	For most protocols $t\ex$ increases only slightly with $L_x$, which does not match the scaling obtained for the bulk correlations (also plotted in \fig{fig:t-XL}).
	This observation disqualifies the density correlations as origin of the nonmonotonic $\Delta \pwall$.

	\section{Discussion and Conclusion}\label{sec:con}
	
	The quenched granular quasi \ac{2d} system considered here exhibits many nontrivial phenomena,
	several of which were observed upon disturbing the steady state by changing the driving amplitude.
	The first is the emergence of propagating density correlations in bulk on length scales beyond the commonly known fluid structure upon heating or moderate cooling \bracket{type~I; \subfigs{fig:corr-bqA}{a}--\sub{d}}.
	
	Qualitatively such an effect has been predicted recently for soft Brownian particles.
	However, we identified several crucial differences regarding the dynamic scaling and the functional form of the correlations.
	An assumption in the previous studies of Brownian systems that is not met by our setup is that of instantaneous temperature change.
	Here we change the driving amplitude instantaneously and the temperature of the inertial particles slowly adapts to the new amplitude.
	This was shown to make an important difference as saturation of the temperature causes a crossover in the scaling behavior (\fig{fig:XS-CS-t}).
	Remarkably, correlation functions were shown to collapse onto universal position-dependent curves (\ie scaling functions) when rescaled appropriately (\fig{fig:C-X-rescaled}).
	This universality is robust across different simulation sizes.
	Although we did not decipher the various features of these curves, we also ascribe their emergence to the gradual temperature change.
	
	In the saturated (\ie long-time) regime we could identify the scaling of the correlation strength $\beta=-2$ as a universal geometric property.
	On the other hand, the ballistic scaling of the dynamic exponent $\alpha=1$ is remarkable because the individual particle motion is diffusive at the considered length scales.
	For an explanation of the constant propagation speed we proposed a semiquantitative mechanism.
	
	Similar studies of a single event in a granular medium at rest yield an exponent $\alpha=1/3$~\cite{jabeen2010universal}.
	The momentum there is solely supplied by the initial event and is split among particles, which decreases the propagation speed in time.
	In contrast, we addressed a medium at nonzero temperature, which is capable of transporting the generated momentum pairs at constant speed.
	Conclusive verification of the proposed semiquantitative mechanism yet remains an open task for future studies.
	
	The present explanation relies on momentum conservation in horizontal directions.
	Therefore it is unclear whether the observed correlations are also present in more sophisticated models or experiments that feature tangential friction.
	Another open puzzle is the determination of the origin of the salient functional forms of the scaling functions.
	A fruitful path to this end may emerge via the testing of a local conservation law with a possible source term via explicit sampling of particle currents~\cite{schindler2016dynamic}.
	
	The asymmetry between energy gain and dissipation induces an asymmetry between heating and cooling.
	Hence, when cooling the fluid starting from a large $A\init$ we observe an additional effect named clustering instability \bracket{type~II; \subfigs{fig:corr-bqA}{e} and \sub{f}}.
	This effect, which is well known for inelastic systems, is greater and hence overlays the former one.
	In particular, the true length scale of the clustering instability could not be determined in our simulations, as it exceeded even our largest simulation boxes.
	
	We further investigated the finite-size scaling of temperature and pressure by adding side walls to the setup.
	While the internal pressure follows the scaling of the temperature, the pressure exerted on the side walls behaves differently.
	In the steady state, the side wall pressure deviation from the infinite-size limit has the opposite sign than the temperature difference (\fig{fig:P-cs}).
	Moreover, after a quench we observe nonmonotonic behavior in the side wall pressure (\fig{fig:P-cq}).
	Numerical data collapse revealed that this is a consequence of an enhanced relaxation speed of the fluid between the side walls (\fig{fig:scaling-cq}).
	In combination with the anomalous steady-state behavior, this results in an over- or undershoot in the side wall pressure that is not observed in the temperature or the internal pressure.
	
	The finite-size scaling $\propto L_x^{-1}$ of temperature and pressure as well as relaxation speed corresponds to the fraction of wall size over system area.
	This is a clear indicator for a change of the bulk properties of the fluid between the walls which constitutes a density induced effect.
	Fluctuation induced wall effects on the other hand, which should exhibit the scaling exponent of the bulk correlations $\beta=-2$, were not detected (\cf\ \fig{fig:t-XL}).
	
	Future plans include investigations of a setup with tuned densities inside and outside the slit such that net side wall pressure in steady state is zero, in order to isolate the transient contribution.
	A further open question and a possible next objective is the study of forces between compact inclusions immersed in the fluid.
	
	\begin{acknowledgments}
		We thank Sebastian Kapfer for fruitful discussions throughout the genesis of this article.
		TS was supported by the Deutsche Forschungsgemeinschaft as part of the Forschergruppe GPSRS under Grant No.\ ME1361/13-2.
	\end{acknowledgments}
	
	\appendix
	
	\section*{Steady-state bulk}\label{sec:steady-bulk}
	
	\begin{figure}
		\includegraphics[width=\halfnl]{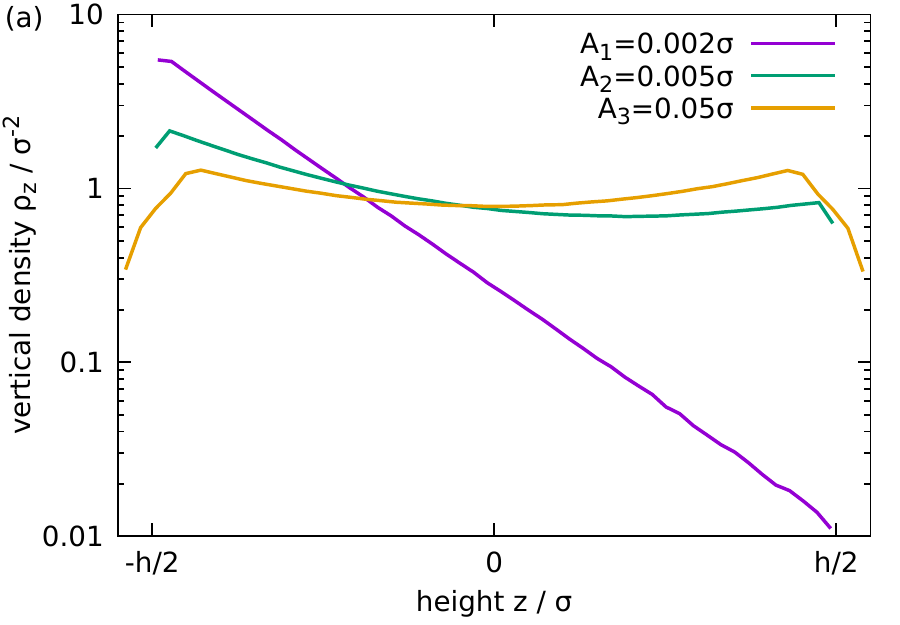}
		\includegraphics[width=\halfnl]{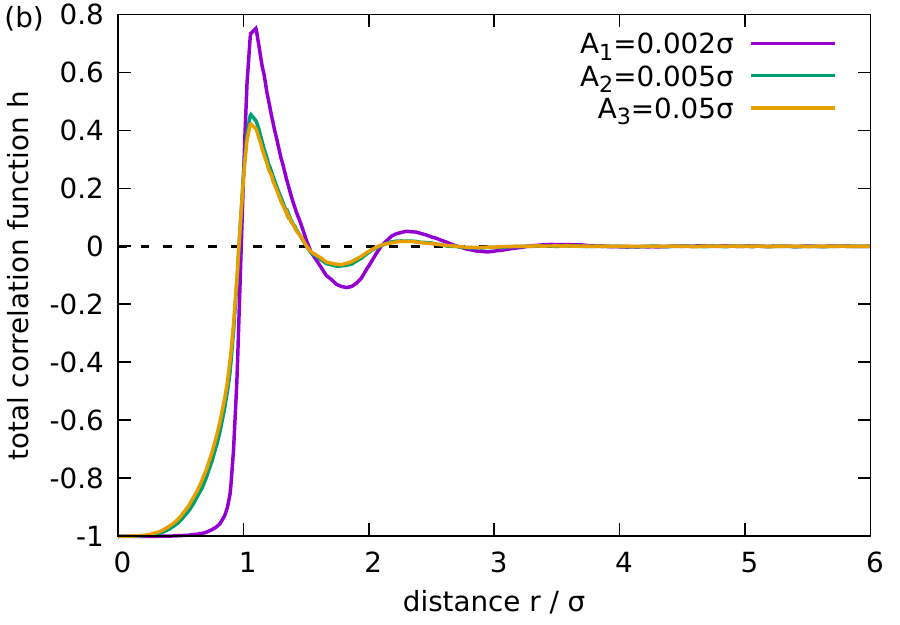}
		\caption{\label{fig:bs}%
			\sub{a} Vertical density distribution $\rho_z$ as a function of the $z$ coordinate, and
			\sub{b} projected total correlation function $h$ as a function of distance $r$,
			for steady-state fluids at amplitudes $A_1$, $A_2$, and $A_3$.
		}
	\end{figure}
	
	Here, we discuss the steady-state fluid as present in the initial state prior to the quench and in the final state infinitely long time after the quench.
	The aim is to demonstrate how stratification creates effectively soft particles.
	\subfigure{fig:bs}{a} shows the vertical density distribution
	\begin{equation}
		\rho_z(z)\equiv \frac{L_z}{L^2}\left\langle\sum_{i=1}^{N}\delta(z_i(t) - z)\right\rangle
	\end{equation}
	at the three investigated $A$ (normalized such that its mean equals $\rho$).
	At high amplitude $A_3$, $\rho_z$ is almost symmetric and the particles fill the whole space between the plates.
	The peaks at the top and bottom plates originate from the mutual repulsion of the particles.
	At low amplitude $A_1$, however, we observe strongly barometric (\ie exponentially decaying) distribution of particles, where most particles are located near the bottom plate and are only slightly hopping.
	At intermediate amplitude $A_2$ the particles are partly stratified.
	
	The differences in stratification also make an impact on $h$ \bracket{shown in \subfig{fig:bs}{b}}.
	As mentioned before, we obtain $h$ from the projected $xy$ coordinates of the particles.
	Therefore, even though the particles are hard and cannot penetrate each other, we can observe a nonzero contribution of $h$ at $r<\sigma$ originating from particles that are (partly) on top of each other.
	This contribution is larger at high amplitudes $A_2$ and $A_3$, where particles fill the whole space between the plates, and smaller at low amplitude $A_1$, where most of the particles populate a single layer near the bottom plate.
	In a \ac{2d} description of the system one can therefore consider the fluid as effectively soft.
	
	\bibliography{literature-paper}
	
\end{document}